\documentclass[amssymb,amsmath,preprint,showkeys,showpacs]{revtex4-1}

\usepackage[utf8]{inputenc} 
\usepackage{amsmath, amssymb, graphicx, subfigure, mathrsfs, xcolor, anysize, setspace}
\usepackage{hyperref}

\begin{document}
\renewcommand{\labelitemi}{$\circ$}
\renewcommand{\labelenumi}{(\alph{enumi})}
\renewcommand{\theequation}{\thesubsection{.\arabic{equation}}}
\renewcommand{\thefigure}{\thesubsection{.\arabic{figure}}}
\renewcommand{\thesubsection}{\arabic{subsection}}
 
\renewcommand{\rq}[1]{
  \setcounter{subsection}{#1}
  \setcounter{equation}{0}
  \setcounter{subsubsection}{0}
  \setcounter{figure}{0}
}

\def \be{\begin{equation}}
\def \ee{\end{equation}}
\def \bea{\begin{eqnarray}}
\def \eea{\end{eqnarray}}

\title{Improved Reissner-Nordstr\"{o}m-(A)dS Black Hole \\in Asymptotic Safety}

\author{Cristopher Gonz\'{a}lez}
\email{cdgonzalez1@uc.cl}
\author{Benjamin Koch}
\email{bkoch@fis.puc.cl}
\affiliation{
Instituto de F\'{i}sica, \\
Pontificia Universidad Cat\'{o}lica de Chile, \\
Av. Vicu\~{n}a Mackenna 4860, \\
Santiago, Chile \\}

\begin{abstract}
This paper studies the quantum modifications of the Reissner-Nordstr\"{o}m-(A)dS black hole within Quantum Einstein Gravity, coupled to an electromagnetic sector.
Quantum effects are introduced on the level of the improvements of the classical solution, where
the originally constant couplings ($G_0$, $\Lambda_0$, and $\alpha_0$)
are promoted to scale dependent quantities ($G_k$, $\Lambda_k$, and $\alpha_k$).
Those running couplings are calculated in the functional renormalization group approach.
A crucial point of this, so called ``improving solutions'' procedure is the scale setting where
the arbitrary scale $k$ acquires physical meaning due to a relation to the coordinate scale $r$.
It is proposed to use such scale settings which are stable after iterative improvements.
Using this method one finds that for those improved solutions, there is no stable remnant
and due to the appearance of a new internal horizon, there is also no necessity
to impose a minimal black hole mass for charged black holes, in order to avoid the the cosmic censorship hypothesis.
\end{abstract}



\maketitle
\tableofcontents

\section{Introduction}\rq{1}

Classical black holes are classical solutions of the equations of general relativity
that have attracted a lot of interest.
The Schwarzschild and the Kerr-Newman black holes
are excellent candidates for numerous astrophysical observations \citep{Celotti:1999tg,Narayan:2013gca}.
Apart from this observational fact, 
black hole solutions are also theoretically extremely interesting objects \citep{Hawking:1969sw}, since
they allow to study the theory of general relativity at its limits of validity: 
The transition from classical to quantum.
The famous Hawking radiation \citep{Hawking:1975,PhysRevD.15.2738} is exemplary for this interplay between classical
geometry and quantum physics.
An other interesting example are charged Reissner-Nordstr\"om black holes,
which via the cosmic censorship hypothesis proposed by Roger Penrose \citep{Penrose:1969pc,Penrose:1973ns,Hawking:1979ig,Penrose:1999vj} allows to derive a minimal mass to charge
ratio for viable black holes. This hypothesis is an open question in theoretical physics and there is no general proof of its validity, however, there are studies that approve and disapprove this conjecture in different scenarios \citep{Wald:1997wa, Brady:1998au, Hod:2008zza}.

It is well known, the formulation of a quantum version of general relativity
is facing serious problems and among the approaches
that try to address this problem, Weinbergs Asymptotic Safety (AS) scenario
is a serious candidate \citep{Hawking:1979ig,Weinberg:1996kw}. 
The key point of this idea is that it conjectures the existence of a non-Gaussian
Fixed Point (NGFP) in the flow of the dimensionless couplings of the gravitational theory \citep{Reuter:1996cp}.
With the techniques of the Functional Renormalization Group (FRG)
one can derive non-perturbative flow equations which allow to study this idea in practice.
Strong evidence for the existence of a non-trivial Ultra Violet (UV) fixed point has been found 
\citep{Reuter:2001ag,lrr-2006-5,Reuter:2007rv,Percacci:2007sz,Litim:2008tt,Reuter:2012id}.
This, is a very important formal result and the question arises whether
and how it is manifest in specific gravitation systems.

Given the importance of black hole physics as testing ground for
quantum gravity candidates, it is clear that within AS and FRG one will
have to say something about black holes. In order to estimate
the leading effects of AS on black holes one can borrow a tool from early quantum electro dynamics.
In this very well tested quantum field theory leading quantum corrections
to the classical Coulomb potential can be obtained by the ``improving solutions'' scheme,
leading to the Uehling potential \cite{Uehling:1935}. Even though 
for next to leading results, other techniques turned out to be more precise, the leading
effects can still be cast within the ``improving solutions'' language.
This technique can be straight forwardly adapted to classical black hole solutions in general relativity.
There exist numerous studies on this subject in terms of using FRG 
results in order to determine quantum corrections to the classical solution 
\citep{Bonanno:1998ye,Bonanno:2000ep,Bonanno:2006eu,Reuter:2006rg,Reuter:2010xb,Koch:2013owa,Koch:2014cqa,Koch:2010nn},
most prominently the Schwarzschild, the (Anti)-de Sitter ((A)dS), and the Kerr black hole.
This work represents a logical continuation of those studies
in the sense that it explores the AS-FRG effects on the Reissner-Nordstr\"om (RN) solution (and (A)dS-RN black hole),
where particular attention is dedicated to the cosmic censorship.
\newpage
This paper is organized as follows:
In section \ref{classRN} the most
relevant classical properties of the Reissner-Nordstr\"om solution are summarized.
In section \ref{RN-QED} the RG flow of Quantum Einstein Gravity (QEG) coupled to Quantum Electrodynamics (QED)
is given. Connecting the formal FRG results to physical results in the context
of black hole physics involves a scale setting procedure. This is given in section \ref{ScaleSet}.
The FRG improvement of the Reissner-Nordstr\"om solution, based on this scale setting,
is presented in section \ref{SecResults}.
Within this, the improved line element, a new iterative improvement and comparison of alternative improvement schemes, the modified horizon structure,
the cosmic censorship, improved temperature and the modified mass and charge of the black hole are discussed. 
Finally,  the results are summarized and commented with concluding remarks
in section \ref{SecConcl}.

\section{Classical RN Black Holes}\label{classRN}
\begin{figure}[ht!]
  \centering
  \subfigure[\footnotesize{ }]{
    \includegraphics[width=.42\textwidth]{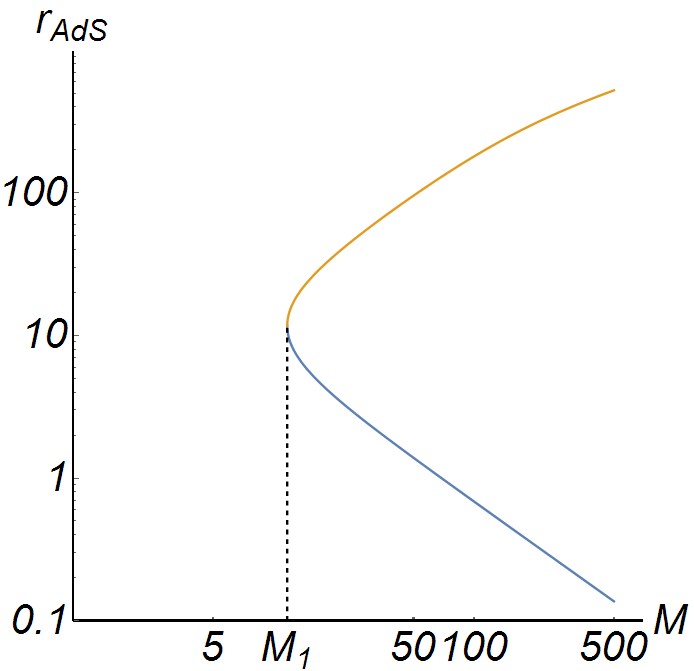}
  }\hspace{1cm}
  \subfigure[\footnotesize{ }]{
    \includegraphics[width=.42\textwidth]{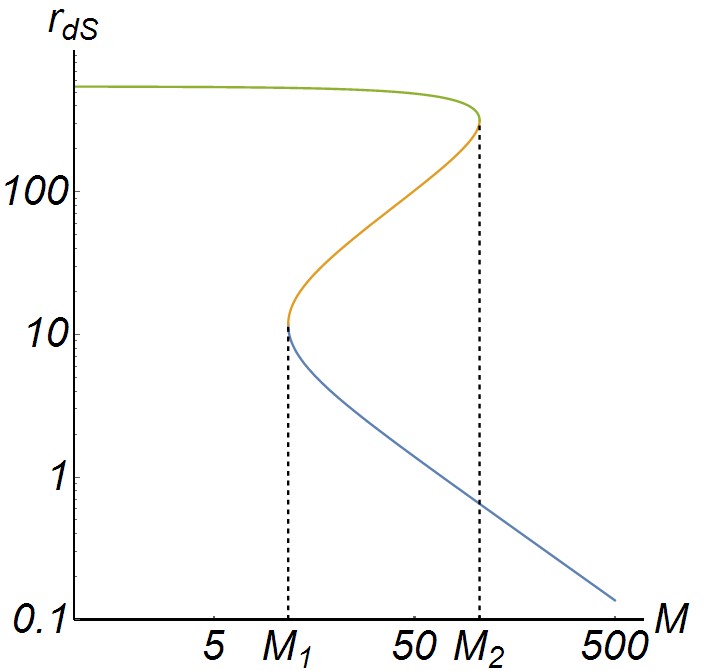}
  }
\caption{\footnotesize{(a) Horizons for the AdS case as a function of the mass parameter $M$. The blue line corresponds
to the inner horizon  $\tilde{r}_1$ and the yellow line to the outer horizon  $\tilde{r}_2$. The merging occurs at $M_{1}\approx 11.7$.\\
(b) Horizons for the dS case as a function of the mass parameter $M$. The blue line corresponds to the inner horizon $r_1$, the
yellow line to the outer horizon $r_2$ and the green line to the cosmological horizon $r_4$.
The outer and the inner horizons merge at mass values very close to the AdS case $M_{1}\approx 11.7$. In addition to this
the outer and the cosmological horizon merge at a mass parameter $M_{2}\approx 105.6$.
The remaining parameters were chosen as  $G_0=1, Q_0=1, \alpha_0=1/137, \Lambda_0=\pm 10^{-5}$.}
}
\label{fig:horizons1}
\end{figure}

\rq{2}
The classical solution for a charged spherically symmetric black hole with cosmological constant is
given by the line element \citep{PhysRevD.19.421,PhysRevD.41.403,Romans:1991nq}
\begin{equation}
\label{metric}ds^2=-f(r)dt^2+\frac{dr^2}{f(r)}+r^2d\Omega^2~,
\end{equation}
where
\begin{equation}\label{fr}
f_{cl}(r)=1-\frac{2G_0 M_0}{r}+\frac{G_0 Q_0^2}{\alpha_0 r^2}-\frac{1}{3}\Lambda_0 r^2~.
\end{equation}
Like in most of the literature the units are chosen such that $c=\hbar=1$, whereas the
electromagnetic fine structure constant $\alpha_0\equiv\frac{e^2}{4\pi}$ is kept explicit.
It is well know this solution collapses for small values of the radial coordinate,
which can be seen from evaluating invariant quantities such as the Kretschmann scalar
in this limit
\begin{equation}
R_{\mu\nu\rho\sigma}R^{\mu\nu\rho\sigma}=\frac{48G_0^2M_0^2}{r^6}+\frac{56G_0^2Q_0^4}{\alpha_0^2r^8}-\frac{96G_0^2M_0Q_0^2}{\alpha_0 r^7}+\frac{8}{3}\Lambda_0^2~,
\end{equation}
One observes that the first two leading divergencies
are dominated by the charge parameter $Q_0$ and that the subleading $1/r^6$ divergence
is the one known from the Schwarzschild solution.
The possible horizons are given by the zeros of the function \eqref{fr} \newpage
\begin{eqnarray}
\label{r1}r_1&=~~~&\rho^{1/2}-\left[\frac{3}{2\Lambda_0}-\rho-\frac{3G_0M_0}{2\Lambda_0}\rho^{-1/2}\right]^{1/2},\\
\label{r2}r_2&=~~~&\rho^{1/2}+\left[\frac{3}{2\Lambda_0}-\rho-\frac{3G_0M_0}{2\Lambda_0}\rho^{-1/2}\right]^{1/2},\\
\label{r3}r_3&=-&\rho^{1/2}-\left[\frac{3}{2\Lambda_0}-\rho+\frac{3G_0M_0}{2\Lambda_0}\rho^{-1/2}\right]^{1/2},\\
\label{r4}r_4&=-&\rho^{1/2}+\left[\frac{3}{2\Lambda_0}-\rho+\frac{3G_0M_0}{2\Lambda_0}\rho^{-1/2}\right]^{1/2},
\end{eqnarray}
where
\begin{eqnarray}
\rho &=&\frac{1}{2\Lambda_0}\left[1-\frac{{\cal R}^{-1/3}{\cal R}_2}{2\alpha_0}-\frac{{\cal R}^{1/3}}{2\alpha_0}\right],\\
\label{R}{\cal R}&=&{\cal R}_1+\sqrt{{{\cal R}_1}^2-{{\cal R}_2}^3,}
\end{eqnarray}
\begin{eqnarray}
{\cal R}_1&=&\alpha_0^3+12G_0Q_0^2\alpha_0^2\Lambda_0-18G_0^2M_0^2\alpha_0^3\Lambda_0,\\
{\cal R}_2&=&\alpha_0^2-4G_0Q_0^2\alpha_0\Lambda_0.
\end{eqnarray}
Out of those four candidates, only those correspond to physical horizons that are real and positive valued.
In order to have at least one physical horizon it is necessary that ${\cal R}_1^2-{\cal R}_2^3\ge0$ in \eqref{R}.
This implies that there exists a minimal value for the mass parameter $M_0$, associated with a critical black hole
with at least one degenerate horizon. This critical value is obtained by writing out the (in)equality explicitly
\begin{equation*}
81G_0^3\alpha_0^3\Lambda_0 M_0^4-(9 G_0 \alpha_0^3+108 G_0^2 Q_0^2 \alpha_0^2 \Lambda_0)M_0^2+24 G_0 Q_0^4 \alpha_0 \Lambda_0 + 16 G_0^2 Q_0^6 \Lambda_0^2 + 9 Q_0^2 \alpha_0^2=0,
\end{equation*}
and solving for  the mass parameter $M_0$.
The two positive valued solutions are
\begin{equation}\label{M1}
M_1=\frac{1}{3G_0}\sqrt{\frac{6G_0Q_0^2}{\alpha_0}+\frac{1}{2\Lambda_0}\left[1-\left(1-\frac{4G_0Q_0^2\Lambda_0}{\alpha_0}\right)^{3/2}\right]}
\end{equation}
\begin{equation}\label{M2}
M_2=\frac{1}{3G_0}\sqrt{\frac{6G_0Q_0^2}{\alpha_0}+\frac{1}{2\Lambda_0}\left[1+\left(1-\frac{4G_0Q_0^2\Lambda_0}{\alpha_0}\right)^{3/2}\right]}~.
\end{equation}
For the case of negative cosmolgical constant $\Lambda_0<0$, there is only one critical mass  $\tilde{M}\equiv M_1$ for which
the cosmological and the Schwarzschild horizon merge. 
For $M>\tilde M$ one finds two physical horizons at
\begin{eqnarray}
\tilde{r}_1&\equiv &r_1,\\
\tilde{r}_2&\equiv &r_2\quad,
\end{eqnarray}
whereas for $M< \tilde M$ the physical horizons disappear and one finds a naked singularity.
This behavior is shown on the left hand side of figure \ref{fig:horizons1}.
For the de Sitter case ($\Lambda_0>0$) a cosmological horizon appears, given by
\begin{equation}
\label{rc}r_{c}=\sqrt{\frac{3}{2\Lambda_0}}\left(1+\sqrt{1+\frac{4G_0Q_0^2\Lambda_0}{3\alpha_0}}\right)^{1/2}=r_4|_{M=0}\quad,
\end{equation}
which for $Q_0\rightarrow0$ recovers the cosmological horizon of the Schwarzschild-de Sitter case.
On the other hand, for $\Lambda_0\rightarrow 0$ the cosmological horizon \eqref{rc} vanishes and the horizons $r_1$ and $r_2$
turn out to be identical to the Reissner-Nordstr\"om case.
Thus, the existence of an intermediate horizon $r_2$ in dS is limited to a maximal and a minimal mass parameter $M$
as it can also be seen from figure \ref{fig:horizons1}b.
Among those horizons, the external horizon \eqref{r2} is of special interest,
since it allows to calculate the Hawking temperature \citep{Hawking:1975,PhysRevD.15.2738} by
\begin{equation}\label{class_temp}
T=\frac{1}{4\pi}f'(r)|_{r=r_2}\quad.
\end{equation}
Even though this concept of temperature has been derived for black 
hole actions that are solutions of the classical field equations, it will be
also applied to the improved solutions, which have the same asymptotic behavior
at infinity.

\begin{figure}[ht!]
  \centering
    \includegraphics[width=.44\textwidth]{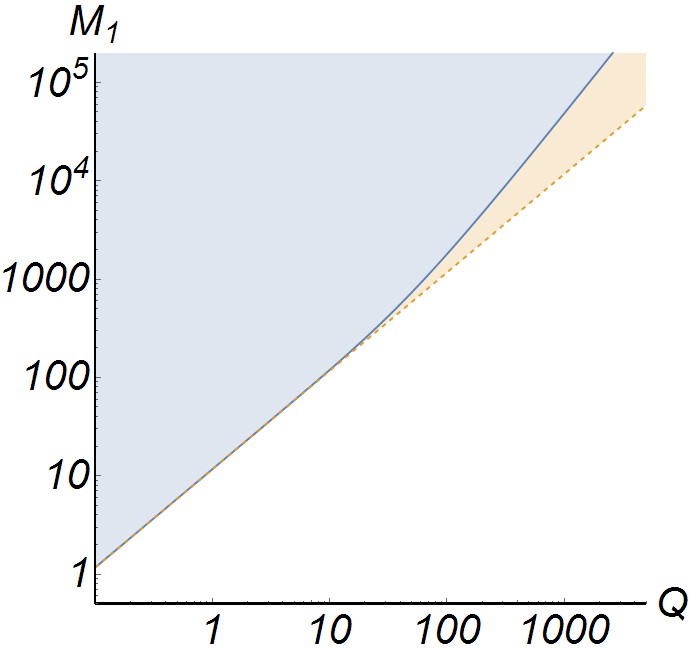}
\caption{\footnotesize{Cosmic censorship scenario, where we plot the equation (\ref{M1}) in the AdS case. The blue line indicate the minimal mass values for the black hole has horizons and the blue zone represents the set of parameters that ensures weak cosmic censorship. The dashed orange line is the 
Reissner Nordstr\"om limit $Q_0/\sqrt{G_0\alpha_0}$. We use the values $\alpha_0=1/137,~G_0=1$ and $\Lambda_0=-10^{-5}$.}}
\label{fig:Cosmic}
\end{figure}

An other particularly interesting concept for charged black holes 
is the cosmic censorship hypothesis. 
This has two forms, the weak and strong censorship. The weak states that that in a process of gravitational collapse, the space-time singularities are hidden by event horizons.  
In other words, one expect only to find black holes that have horizons, such that the singularity is ``dressed" by the horizon \citep{Penrose:1973ns}.  
For the solutions presented here, cosmic censorship condition implies that only those black holes are physical
that have at least one real valued inner horizon. This condition is solved by equation \eqref{M1},
which, for given parameters ($\Lambda_0,\,Q_0\, \dots$), implies a minimal physical mass.
In figure \ref{fig:Cosmic} it is illustrated, how this minimal physical mass depends on the charge of a given (A)dS solution.
One observes that for $\Lambda_0\neq 0$, simple linear relation of the Reissner-Nordstr\"om censorship 
\be\label{RNcens}
M_{crit}|_{\Lambda_0=0}=Q_0/\sqrt{G_0\alpha_0}\quad,
\ee
gets modified for larger values of $Q_0$.

\section{Renormalization group flows in quantum Einstein gravity coupled to QED}
\label{RN-QED}
\rq{3}

\begin{figure}[ht!]
  \centering
  \subfigure[]{
    \includegraphics[width=.44\textwidth]{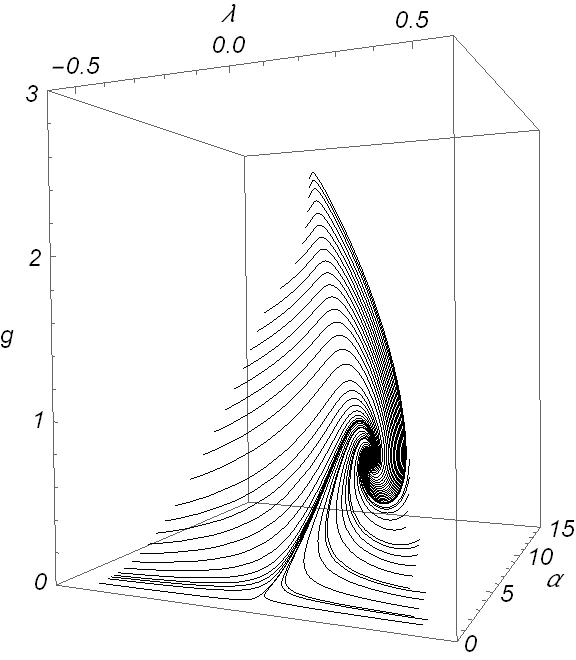}
  }\hspace{1cm}
  \subfigure[]{
    \includegraphics[width=.44\textwidth]{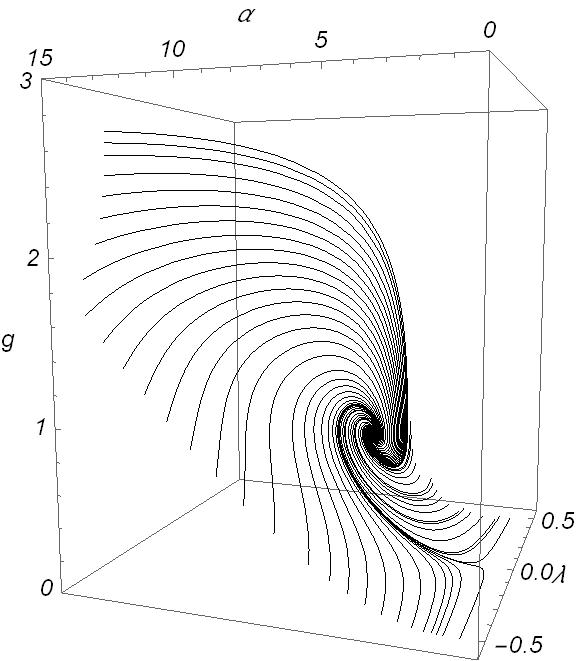}
  }
  \caption{RG flow of the three couplings $g(k)$, $\lambda(k)$, and $\alpha(k)$ 
  that solutions of \eqref{RGFlow} with the non-trivial fixed point \eqref{fpval}.
  \label{fig:flow}}
\end{figure}

In the Einstein-Hilbert truncation coupled to an electromagnetic sector, the effective action is given by \citep{Daum:2009dn,Harst:2011zx}
\begin{eqnarray}\label{EHMact}
\nonumber\Gamma_{k}&=&\Gamma_{k}^{\text{grav}}+\Gamma_{k}^{\text{``QED''}}\\
&=&\frac{1}{16\pi G_k}\int d^4x\sqrt{g}[-R+2\Lambda_k]-\frac{1}{4\alpha_{k}}\int d^4x\sqrt{g}F_{\mu\nu}F^{\mu\nu}\quad,
\end{eqnarray}
where the three coupling constants are, the Newton coupling $G_k$, the cosmological coupling $\Lambda_k$,
and the electromagnetic coupling $\alpha_k$ (in terms of the fine structure ``constant'').
The scale dependence of those couplings is indicated by the subindex $k$ which has energy dimension one.
The dimensionless couplings are obtained from the dimensionfull couplings by multiplying with the corresponding
power of $k$
\begin{equation}\label{dimensionless}
g_k=G_k k^2, \hspace{1cm} \lambda_k=\Lambda_k k^{-2}, \hspace{1cm} \alpha_k=\alpha_k\quad.
\end{equation}
The evolution of the dimensionless couplings \eqref{dimensionless} is governed by the renormalization group equations \citep{Reuter:1996cp,Harst:2011zx,Reuter:2001ag}.
\begin{equation}\label{RGFlow}
k\partial_k g_k=\beta_g(g_k,\lambda_k)\;, \hspace{1cm} k\partial_k \lambda_k=\beta_\lambda(g_k,\lambda_k)\;, 
\hspace{1cm} k\partial_k \alpha_k=\beta_{\alpha}(g_k,\alpha_k)\;.
\end{equation}
The beta functions corresponding to $g$ and $\lambda$ are calculated in \citep{Reuter:1996cp} 
and the beta function for $\alpha$ is obtained in \citep{Daum:2009dn} for the general case of Einstein-Hilbert action coupled to a 
Yang-Mills field and adaptated in \citep{Harst:2011zx} 
for the QEG picture coupled to QED. All these $\beta$-fuctions are computed in a $d$-dimensional 
spacetime, independent of the curvature of the gravitational background.  For $d=4$ one obtains\newpage
\begin{eqnarray}
\label{beta_l}\beta_\lambda(g,\lambda)&=&(\eta_N-2)\lambda+\frac{1}{2\pi}g\left[10\Phi_2^1(-2\lambda)-8\Phi_2^1(0)-5\tilde{\Phi}_2^1(0)\right],\\
\label{beta_g}\beta_g(g,\lambda)&=&(2+\eta_N)g\;,\\
\label{beta_a}\beta_\alpha(g,\alpha)&\equiv&\left(Ah_2(\alpha)-\frac{6}{\pi}\Phi_1^1(0)g\right)\alpha\quad.
\end{eqnarray}
The anomalous dimension of the gravitation coupling is given by
\begin{equation}
\eta_N(g,\lambda)=\frac{gB_1(\lambda)}{1-gB_2(\lambda)}\quad,
\end{equation}
where the two functions of the adimensional constant $\lambda$ are given by
\begin{eqnarray}
B_1(\lambda)&\equiv &\frac{1}{3\pi}\left[5\Phi_1^1(-2\lambda)-18\Phi_2^2(-2\lambda)-4\Phi_1^1(0)-6\Phi_2^2(0)\right],\\ 
B_2(\lambda)&\equiv &-\frac{1}{6\pi}\left[5\tilde{\Phi}_1^1(-2\lambda)-18\tilde{\Phi}_2^2(-2\lambda)\right]\quad.
\end{eqnarray}
The functions $\Phi_i$ have been calculated in the ``optimised cutoff" sheme \citep{Litim:2001up,Litim:2003vp}
\begin{equation}\label{Phi}
\Phi_n^p(w)=\frac{1}{\Gamma(n+1)}\frac{1}{(1+w)^p}\;,\hspace{1cm}\tilde{\Phi}_n^p(w)=\frac{1}{\Gamma(n+2)}\frac{1}{(1+w)^p}\quad.
\end{equation}
Finally, the constant $A$ and the function $h_2(\alpha)$ in equation \eqref{beta_a} are defined in \citep{Harst:2011zx}.\\
The RG equations \eqref{RGFlow} can be solved numerically, and depicted as a three dimensional flow graphic \ref{fig:flow}.
One observes, the existence of a non trivial fixed point at which $\beta_{g}=0,~\beta_{\lambda}=0~\text{and}~\beta_{\alpha}=0$.
This fixed point is located at the coupling values
\begin{equation}\label{fpval}
g_{*}=0.707\;, \hspace{1cm} \lambda_{*}=0.193\;, \hspace{1cm} \alpha_{*}=6.365\quad
\end{equation}
and it is approached in the UV by specific set of trajectories. 
By using \eqref{fpval} the dimensionfull coupling constants can be approximated 
at the vicinity of this fixed point
\begin{equation}\label{UV}
\lim_{k\rightarrow\infty} G_k=g_{*}k^{-2}, \hspace{.5cm}\lim_{k\rightarrow\infty} \Lambda_k=\lambda_{*}k^{2}, \hspace{.5cm}\lim_{k\rightarrow\infty} \alpha_k=\alpha_{*}\quad.
\end{equation}
\newpage
It is sometimes convenient to work with analytic approximations of the renomalization group flow \eqref{RGFlow} 
in this case the following approximation will be used \citep{Harst:2011zx,Contreras:2013hua,Koch:2013rwa,Koch:2010nn}
\begin{eqnarray}
\label{g(k)}g(k)&=&\frac{G_0 k^2}{1+\frac{G_0}{g_{*}}(k^2-k_0^2)}\,,\\
\label{lambda(k)}\lambda(k)&=&\frac{\Lambda_0}{k^2}+\lambda_{*}\left(1-\frac{k_0^2}{k^2}\right)+\frac{g_{*}\lambda_{*}}{G_0 k^2}\log{\left(\frac{1+\frac{G_0}{g_{*}}k_0^2}{1+\frac{G_0}{g_{*}}k^2}\right)},
\end{eqnarray}
\begin{eqnarray}
\nonumber \alpha(k)^{-1}&=&\left[1+\frac{G_0}{g_{*}}(k^2-k_0^2)\right]^{\frac{3\Phi g_{*}}{\pi}}\left[\frac{1}{\alpha_0}-\frac{g_{*}}{\alpha_{*} G_0 k_0^2}~_2F_{1}\left(1,1,1+\frac{3\Phi g_{*}}{\pi};1-\frac{g_{*}}{G_0k_0^2}\right)\right]\\
\label{alpha(k)}&&+\frac{g_{*}}{\alpha_{*} G_0 k^2}\left[1+\frac{G_0}{g_{*}}(k^2-k_0^2)\right]~_2F_{1}\left(1,1,1+\frac{3\Phi g_{*}}{\pi};\frac{G_0k_0^2-g_{*}}{G_0k^2}\right)\quad,
\end{eqnarray}
where $_2F_1$ is the hypergeometric function and $\Phi\equiv \Phi_1^1(0)=1$ is evaluated using \eqref{Phi}.
Apart from the closed analytic form those approximated functions have the
advantage that they have a well defined infra-red limit $k\rightarrow k_0$ as it can be seen from taking
the limit with
$$g(k)\rightarrow G_0 k_0^2,~\lambda(k)\rightarrow \Lambda_0/k_0^2,~\alpha(k)\rightarrow \alpha_0\quad.$$

\section{Scale setting for classical backgrounds}
\label{ScaleSet}\rq{4}

\begin{figure}[ht!]
  \centering
  \subfigure[\footnotesize{ }]{
    \includegraphics[width=.42\textwidth]{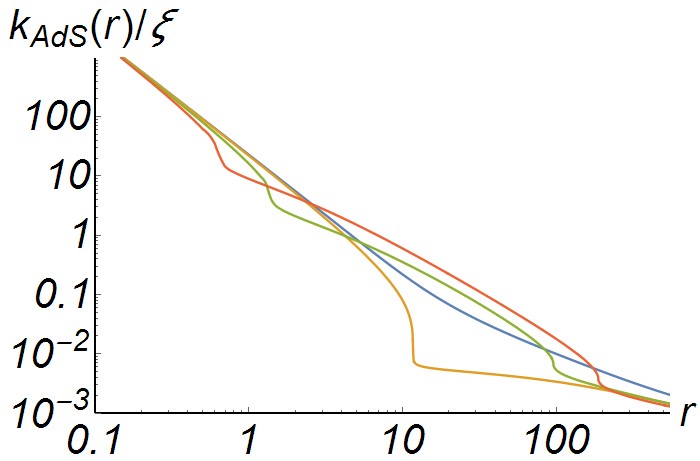}
  }\hspace{1cm}
  \subfigure[\footnotesize{ }]{
    \includegraphics[width=.42\textwidth]{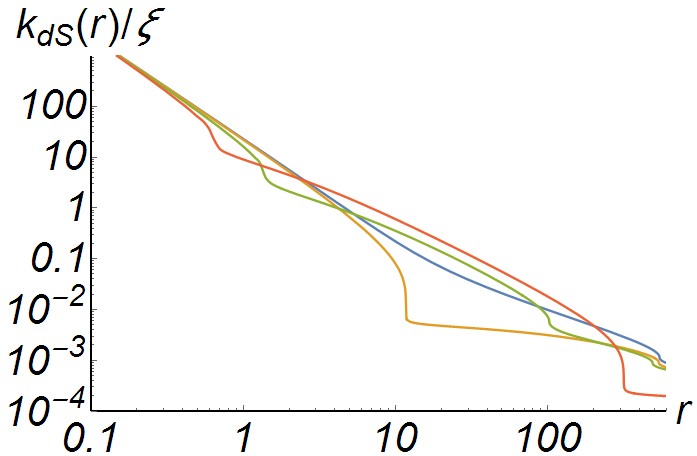}
  }
\caption{\footnotesize{Dependence of the scale $k(r)/\xi$ according to \eqref{k(r)}
for $G_0=1, Q_0=1, \alpha_0=1/137, \Lambda_0=\pm 10^{-5}$.
\newline(a) AdS line element with masses  $M_0$=\{5, 11.7, 50, 105.6\} (blue, yellow, green, and red).
\newline (b) dS line element with the same masses and color codings.}}
\label{fig:kr}
\end{figure}

Due to the renormalization program the coupling constants become scale dependent quantities.
In the context of the Einstein -Hilbert - Maxwell action \eqref{EHMact}, this means
\begin{equation}
G\rightarrow G_k,~\Lambda\rightarrow \Lambda_k~\text{and}~\alpha\rightarrow \alpha_k\quad,
\end{equation}
where $k$ is an arbitrary scale of mass dimension one.
If one is interested in a particular physical context, one tries
to set the arbitrary scale in terms of characteristic physical quantities of the system under consideration. 
This step is crucial for the physical interpretation of the running couplings, in particular in extreme situations, 
where the scale dependence can become strong \citep{Bonanno:1998ye,Bonanno:2000ep,Koch:2010nn,Bonanno:2011yx,Koch:2013owa,Koch:2014cqa,2011arXiv1101.4995F,Bonanno:2001xi,Bonanno:2012jy,Hindmarsh:2012rc,Falls:2012nd,Copeland:2013vva,Koch:2014joa,Koch:2015nva,Bonanno:2015fga} .
Throughout this paper we will follow the procedure outlined in \citep{Bonanno:1998ye,Bonanno:2000ep} which relates the arbitrary
energy scale to an inverse distance scale $k\propto 1/d$.
This type of approach is well motivated since in the context of QED it allows to
obtain the well known and well tested Uehling potential \cite{Uehling:1935}.
It is interesting to note that the electromagnetic coupling used in this paper
\eqref{alpha(k)} can be written like 
the usual running coupling of QED $\alpha^{-1}(k)=-A \ln(k)+ c$, where $c=-A\gamma\psi(3\Phi g_{*}/\pi)$, $\gamma$ is the Euler constant, and $\psi$ 
is the Digamma function \cite{Harst:2011zx}. 
For spherically symmetric black holes with mass and charge in a cosmological background,
the physical quantities characterizing the system are $\alpha_0, Q_0, G_0, M_0,\text{~and~} \Lambda_0$, which will be assumed to be determined at
some large radial scale $r_0$. In additional to those quantities there is the radial parameter $r$, which is expected
to play a crucial role in the scale dependence.
Thus, the scales $k$ and $d$ from the relation $k\propto 1/d$
will be functions of those physical quantities defined at large distance
\begin{equation}\label{k(r)}
k(r, \alpha_0, Q_0, G_0, M_0, \Lambda_0)=k(r)\equiv \frac{\xi}{d(P(r),\alpha_0, Q_0, G_0, M_0, \Lambda_0)}\quad,
\end{equation}
where $P(r)$ is the point in space-time that one likes to study and $d(P)$
is a characteristic length scale separating this point from the black hole.
Please note that using the relation~(\ref{k(r)}) one assumes implicitly
that the black hole is imbedded in vacuum.
For example, if the black hole would be surrounded
by some non-negligible matter density $\rho(r)$ this would introduce
a new local energy scale which could replace (\ref{k(r)}) as the dominant infrared cutoff.
The dimensionless parameter $\xi$ in (\ref{k(r)}) controls the importance of scale dependence,
in the sense that zero or small $\xi$ corresponds to zero or weak scale dependence.
Here, $d(r)$ will be identified with the absolute proper radial distance between the center of
the black hole and the point $P$ calculated along a radial curve ${\cal C}_r$ \citep{Bonanno:2000ep}
\begin{equation}
d(P(r))=\int_{{\cal C}_r}\sqrt{|ds^2|}\quad.
\end{equation}
Please note that different choices for this length scale give 
typically very similar results \citep{Koch:2013owa,Koch:2014cqa}. 
For the black hole metric (\ref{metric}), 
this length scale reads
\begin{equation}\label{d(r)}
d(r)=\int_0^r \frac{dr}{\sqrt{|f(r)|}}=\int_{0}^r\frac{dr}{\sqrt{|1-\frac{2G_0 M_0}{r}+\frac{G_0Q_0^2}{\alpha_0 r^2}-\frac{1}{3}\Lambda_0 r^2|}}\quad.
\end{equation}
In figure \ref{fig:kr}, the radial dependence of $k/\xi=1/d(r)$ is shown in the dS and the AdS case, for various different masses.
One observes that the $k(r)$ is monotonically decreasing, with its steepest dependence on the horizons of the classical line element.
The dS scale presents three visible vertical steps, one at the internal black hole horizon (which comes due to the charge contribution to the metric), one at the outer horizon, and another one at the cosmological horizon, the latter is absent in the AdS case. Another important point is that the biggest step (yellow line) is for the extreme black hole, where the inner and outer horizons merge. The same holds for the dS case, where external and cosmological horizons merge (red line). The $\Lambda_0=0$ case is equivalent to the AdS case, because this two have the same horizon structure.
\newline
If one consider the pure Reissner-Nordström solution, we can compute an analytical solution for the proper distance, setting $\Lambda_0=0$ in \eqref{fr} or \eqref{d(r)}, which is given by
\begin{eqnarray}\label{drRN}
\nonumber d_\text{RN}(r)&=&r \sqrt{f_\text{RN}(r)}+G_0 M \log \left|r \sqrt{f_\text{RN}(r)}+r-G_0 M\right|\\
&&+G_0 M \log \left|Q \sqrt{\frac{G_0}{\alpha_0}}\left(1-\frac{M}{Q_0}\sqrt{\alpha_0 G_0}\right) \right|-Q \sqrt{\frac{G_0}{\alpha_0}}\quad,
\end{eqnarray}
where we defines $f_\text{RN}(r)=f(r)|_{\Lambda_0=0}$. This solution only applies for masses different from the critical mass \eqref{RNcens}.

\section{Improved black hole solution in the IR and global behavior}
\label{SecResults}

In one way or the other, the scale dependence of the couplings ($G_k,~\Lambda_k,~\alpha_k$) in the effective action and 
the radial dependence of the actual scale setting $k(r)$, will have to result in a modification of the
actual line element of the quantum corrected black hole space-time.
As first approximation, which is expected to be most reliable in the IR, one can
apply the ``improving the classical solution'' scheme \citep{Bonanno:2000ep,Dittrich:1985yb} , which implements the scale dependence
based on correction of the classical line element \eqref{fr}.

\subsection*{1. Improved line element}\rq{5}

\begin{figure}[ht!]
  \centering
    \includegraphics[width=.5\textwidth]{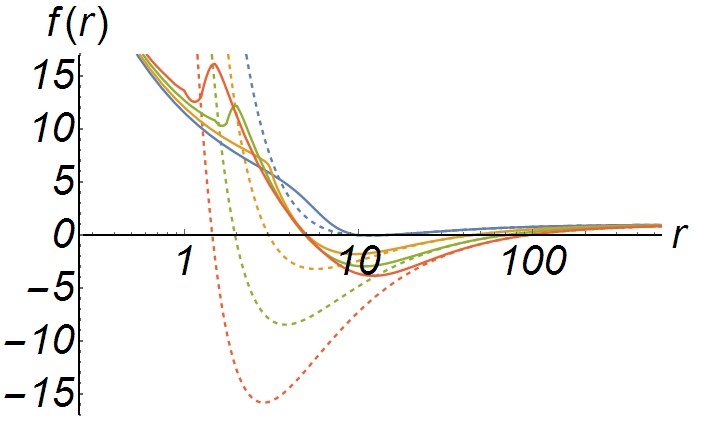}(c)
\caption{\footnotesize{Improved metric function for the self-consistent value $\xi_{sc}$ for: \\
(a) AdS, \\
(b) dS, and \\
(c) $\Lambda_0$=0, with mass values $M_0$=\{12, 24, 36, 48\} (blue, yellow, green, red), and $G_0=1, Q_0=1, \alpha_0=1/137, \Lambda_0=\pm 10^{-5}$, $k_0=0.01$}. The dashed lines are the classical metric function for each case, plotted for comparison.}
\label{fig:fimp}
\end{figure}

In this improvement scheme one promotes the scale independent couplings, that are present in
the classical solution, to the scale dependent quantities known from the RG flow \eqref{beta_l}-\eqref{beta_a}
\begin{equation}\label{f_k}
f_k(r)=1-\frac{2g(k)M_0}{k^2r}+\frac{g(k) Q_0^2}{\alpha(k)k^2 r^2}-\frac{1}{3}\lambda(k)k^2 r^2\quad.
\end{equation}
The arbitrary scale $k$ becomes a physically relevant quantity due to the scale setting \eqref{k(r)} shown in figure \ref{fig:kr}.
With this scale setting one obtains the RG-improved metric function $f(r)$ shown in figure \ref{fig:fimp}. 
For comparison, the purely classical solution is depicted in dashed lines  and the solid lines are the RG-improved $f(r)$. 
Different curves correspond to different parameter choices.
The $\xi$ parameter in \eqref{k(r)} is chosen, using a ``self-consistent'' choice as in \citep{Koch:2013owa}, 
(see also section \ref{sec:impimp}.2)
\begin{equation}\label{xisc}
\xi_{sc}^2\equiv\frac{3}{4\lambda_{*}}\quad,
\end{equation}
In figure \ref{fig:fimp}(a,b) one sees 
that the improved and the classical line elements are very similar for $r\gg M$,
which is of course expected if given that the initial conditions for $\Lambda_0,\, G_0,\, M_0, \alpha_0$,
were obtained experimentally at the large $r$ limit.
On the other hand, in the (A)dS cases, in $r\ll M$ regime, the asymptotic 
behavior $r\rightarrow 0$ of the line element switches, which results in the appearance of 
a new internal horizon.
From figure \ref{fig:fimp}(c) one observes that only for $\Lambda_0=0$, the asymptotic behavior $r\rightarrow 0$ of the classical 
and the improved line element has the same sign and therefore no new horizon appears.
\begin{figure}[ht!]
  \centering
  \subfigure[\footnotesize{ }]{
    \includegraphics[width=.44\textwidth]{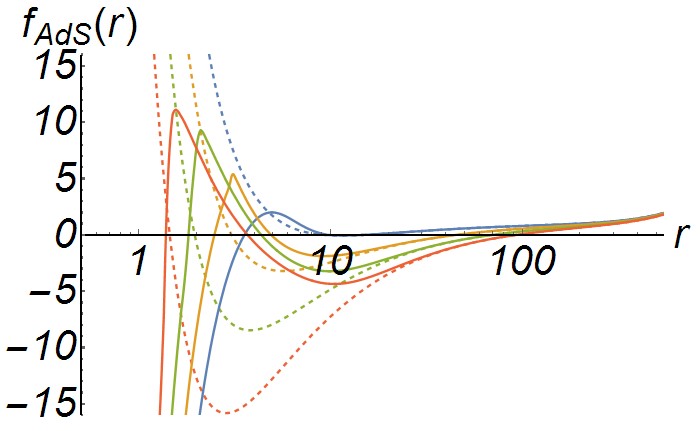}
  }\hspace{.1cm}
  \subfigure[\footnotesize{ }]{
    \includegraphics[width=.44\textwidth]{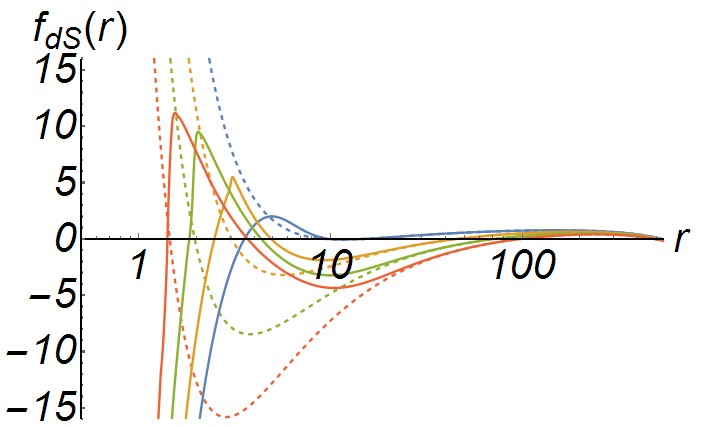}
  }
\end{figure}
\newline
One observes that in all three scenarios in figure \ref{fig:fimp}(a,b,c), the outer black hole
horizon is slightly shifted towards smaller values as with respect to the purely classical solution.

\begin{figure}[ht!]
  \centering
    \includegraphics[width=.6\textwidth]{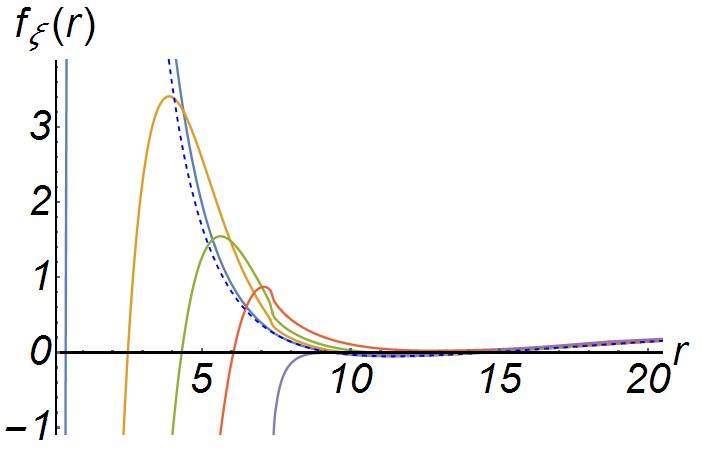}
\caption{\footnotesize{Improved $f(r)$ for AdS case, setting the mass parameter $M_0=12$ and $G_0=1, Q_0=1, \alpha_0=1/137, k_0=0.01$}. We use different $\xi$ values, $\xi=\{0.5,~1.5,~2.5,~4,~11\}$ (blue, yellow, green, red, purple) and the dashed line correspond to the classical solution for comparison with the same parameters. There is
a good agreement of the RG-improvement and the classical solution in the limit $\xi\ll 1$.}
\label{fig:xifr}
\end{figure}

One can further
investigate the dependence of the improved functions for different $\xi$ values. 
From figure \ref{fig:xifr} it can be seen that for larger $\xi$ the difference between the improved
and the classical solution is more pronounced and occurs at larger radial values, where the improvement have a single horizon.
After this general study of the improved line element, one can now turn to more specific 
physical aspects such as horizon structure, cosmic censorship, and temperature.

\subsection*{2. Alternative improvement schemes and improving the improved}\label{sec:impimp}

\begin{figure}[ht!]
  \centering
  \subfigure[\footnotesize{ }]{
    \includegraphics[width=.44\textwidth]{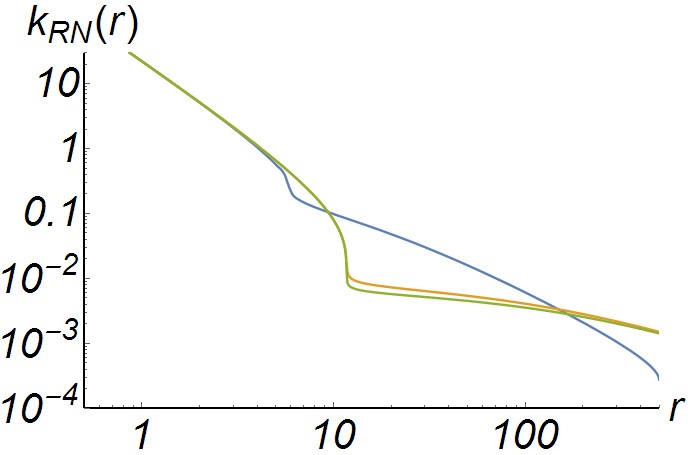}
  }\hspace{1cm}
  \subfigure[\footnotesize{ }]{
    \includegraphics[width=.44\textwidth]{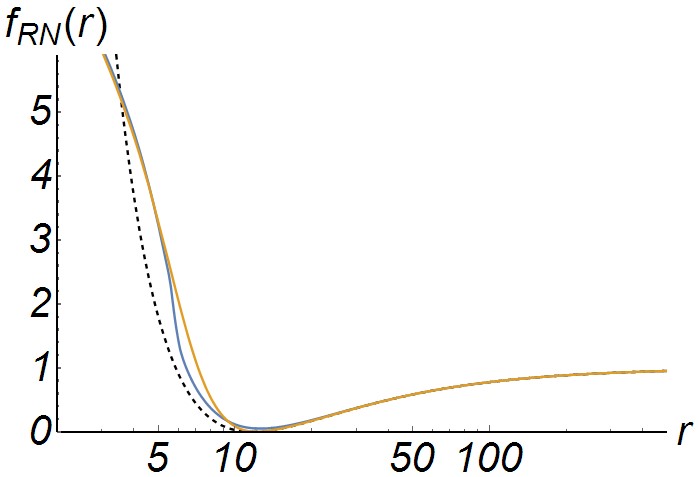}
  }
\caption{\footnotesize{(a) Comparison of the proper time (blue) and numerical distance (orange) and analytical distance (green) scale settings, using $G_0=1, Q_0=1, \alpha_0=1/137$ and a mass value $M\simeq M_{crit}|_{\Lambda_0=0}$ 
(The analytical calculation is plotted only to compare with the numerical integration). One can see that both settings are basically the same until $r\sim 5$. 
Above that, one can see differences between the curves. 
(b) $f(r)$ improved functions for the two scales settings. We use the same values for the constants and $\xi_{sc}$. One can see not much differences between the 2 scale settings at a improved solution level.}}
\label{fig:ScaleCompRN}
\end{figure}

\begin{figure}[ht!]
  \centering
  \subfigure[\footnotesize{ }]{
    \includegraphics[width=.44\textwidth]{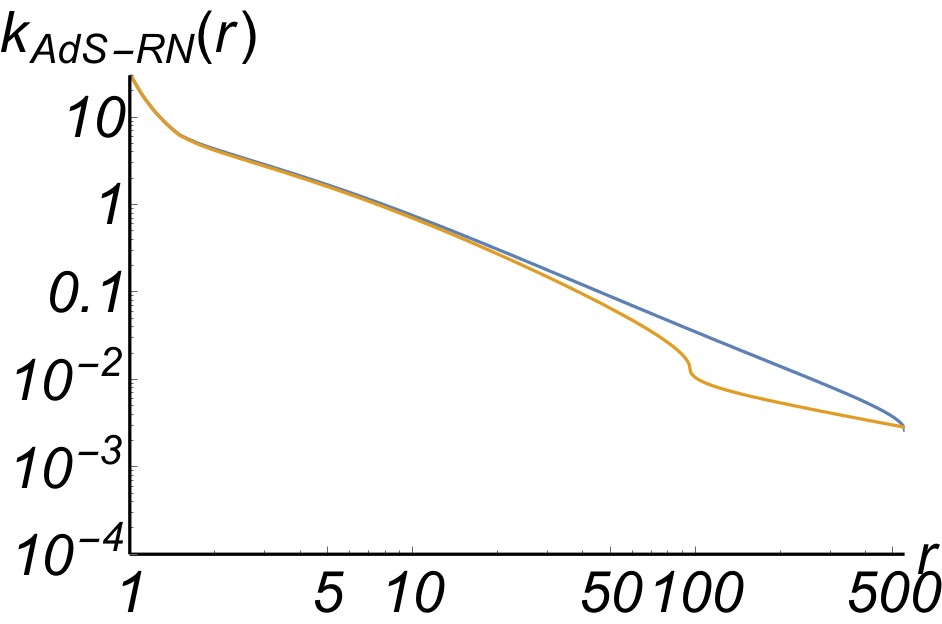}
  }\hspace{1cm}
  \subfigure[\footnotesize{ }]{
    \includegraphics[width=.44\textwidth]{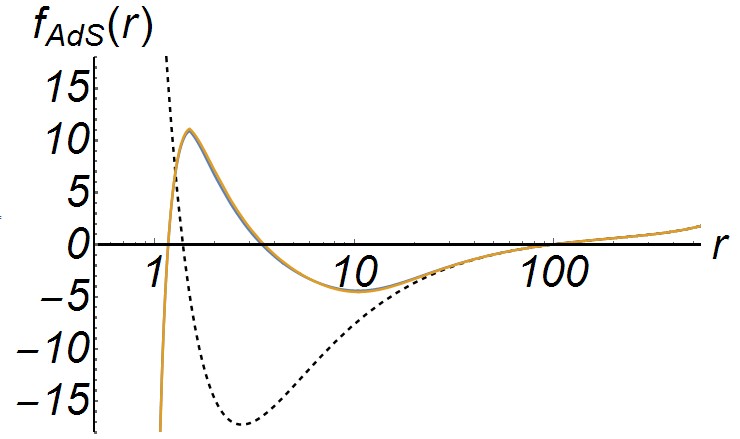}
  }\caption{\footnotesize{
  (a) Comparison of the proper time (blue) and distance (orange) scale setting. 
  (b) Comparison of the improved metric function \eqref{f_k} using proper time (blue) and proper distance (orange). 
  Also we plot the classical solution, in black dashed, for reference. For both plots we are using $\xi_{sc}$ and $G_0=1, Q_0=1, \alpha_0=1/137, \Lambda_0=- 10^{-5}$, $k_0=0.01$ and $M=50$. For the de Sitter case, there is no difference between scale settings and so with the improved functions.
}}
\label{fig:ScaleCompAdSRN}
\end{figure}

There is no rigid principle imposing the scale setting prescription.
Still, as long as one chooses a physically reasonable quantity for the renormalization scale $k$
one can expect that the ``improving solutions scheme'' 
gives a better description of the system under consideration. This works for example for the Uehling potential \cite{Uehling:1935}.
In the case of improved black hole solution an alternative choice for the renormalization scale $k$
would be for example in terms of the proper time rather than in terms of the proper distance.
The length scale associated to the proper time is
\be\label{PropTime}
k(r)=\frac{\xi}{\tau(r)}=\xi\left(\int_0^{r}dr'\left(f(r)-f(r')\right)^{-1/2}\right)^{-1}\quad.
\ee
One can compute this scale numerically and compare it with the corresponding proper distance, numerical or analytical \eqref{drRN}.
For the Reissner Nordstrom case this is done in figure \ref{fig:ScaleCompRN}.
One observes that the choice of the scale setting has no important effect on the form of the improved metric.
\begin{equation}\label{tauRN}
\tau_{RN}(r)=\int_0^{r}dr'\left(\frac{2G_0M}{r'}-\frac{G_0Q^2}{\alpha_0r'^2}-\frac{2G_0M}{r}+\frac{G_0Q^2}{\alpha_0r^2}\right)^{-1/2}\quad.
\end{equation}
In order to see to which extend such a different choice would affect the results of the previous section in the AdS case
we show in figure \ref{fig:ScaleCompAdSRN}(a) the radial dependence of the $k(r)$.
One sees that difference between both scale settings are rather moderate.
In figure \ref{fig:ScaleCompAdSRN}(b) it is further shown how this difference reflects in the improved metric function.
Luckily, one can see that no qualitative changes appear and that quantitative changes
are moderate.
Such, an insensitivity to different scale settings is something which is desirable.

When doing such improvements one would hope that the improvement gets one as close to
the actual answer as possible. One encouraging finding in this sense would be for example
that iterating the improvement procedure converges to a stable line element.
Which choice for $\xi$ in \eqref{k(r)} is optimal in this sense?
In order to get an analytic expression for $\xi$, 
one can calculate \eqref{d(r)} in the high energy limit ($r\ll 1$) 
\begin{equation}\label{d(r)UV0}
d(r)\simeq \frac{1}{2}\sqrt{\frac{\alpha_0}{G_0 Q_0^2}}r^2\left[1+\frac{2}{3}\frac{M_0\alpha_0^2}{Q_0^2}r+{\cal O}(r^2)\right].
\end{equation}
(this procedure is shown in detail in Appendix).
Once this is done, one can get $f(r)$ in UV limit, which is defined as $f_{*}(r)$ (\ref{f*}). 
With this new function, one can calculate a new $d(r)$, given by
\begin{equation}\label{d*0}
d_{*}(r)\simeq \frac{\sqrt{3}}{4}\frac{1}{Q_0\xi}\sqrt{\frac{\alpha_0}{G_0\lambda_{*}}}r^2.
\end{equation}
Comparing the first order improvement \eqref{d(r)UV0} and 
the second order improvement \eqref{d*0}, one finds that both length scales agree,
indicating a (UV) convergence of the improvements if one chooses
\begin{equation}
\xi^2=\xi_{sc}^2\equiv\frac{3}{4\lambda_{*}}\quad.
\end{equation}
This UV-stable choice is similar to the choice which was in previous studies called ``self consistent'' \cite{Koch:2013owa,Koch:2014cqa}.

\begin{figure}[ht!]
\centering
\includegraphics[width=.6\textwidth]{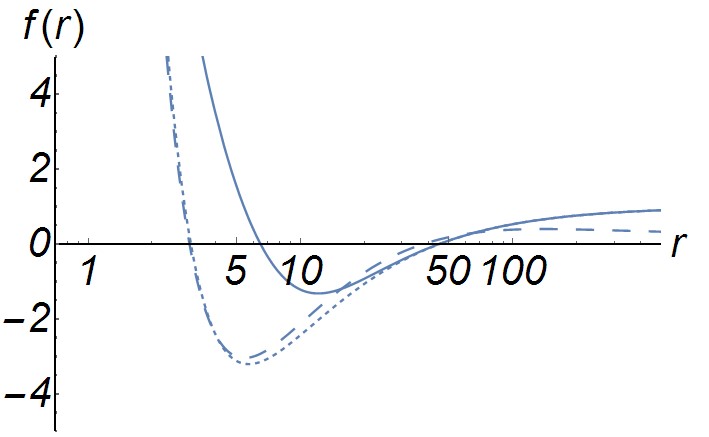}
\caption{\footnotesize{Comparison of $\tilde f(r)$ (dashed), $f_{imp}(r)$ (line) and $f_{cl}(r)$ (point) for the Reissner-Nordstöm black hole, using $\xi_{sc}$, $\epsilon=0.01$, $G_0=1, Q_0=1, \alpha_0=1/137, \Lambda_0=- 10^{-5}$, $k_0=0.01$ and $M=15$}}
\label{fig:ImpActImpSol}
\end{figure}

The framework of modern quantum field theory offers tools to study specific physical systems 
that go beyond the ``improving solutions'' approach. In principle one can for example
derive the quantum ``effective action''. The equations of motion for this effective
action are known as ``gap equations'' and they are typically non-local higher order differential
equations, which are very difficult to solve. 
Still, when restricting to the leading local operators, those equations reduce to second order  equations \cite{Reuter:2003ca}.
By assuming spherical symmetry and that the matter part of the stress energy tensor vanishes exactly for all $r\neq 0$,
one can sometimes solve those gap equations with a Schwarzschild ansatz $g_{00}=-1/g_{11}\equiv \tilde f(r)$. 
Such solutions have been found for the Einstein-Maxwell system and for the Einstein-Hilbert system \cite{Koch:2015nva}.
It is interesting to compare the behavior of the metric functions 
obtained from the improving solutions approach $f_{imp}(r)$ and the metric
function that solves exactly the simplified version of the gap equations $\tilde f(r)$.
One finds that both quantum improved descriptions have a well defined classical limit
\bea
\lim_{\xi \rightarrow 0} f_{imp}(r)&=&f_{cl} (r),\\ \nonumber
\lim_{\epsilon \rightarrow 0 } \tilde f (r) &=&f_{cl}(r). 
\eea
A comparison of both functions is shown in figure \ref{fig:ImpActImpSol} for 
\begin{equation}\label{fImpAct}
\tilde f(r)=
\frac{r^4\epsilon^2 \alpha_0+4\epsilon r^3 \alpha_0+4(1-G_0 M_0\epsilon)r^2 \alpha_0-8rG_0 M_0 \alpha_0+4 G_0 Q_0^2}{4r^2(\epsilon r+1)^2 \alpha_0}
\end{equation}
One observes that $ f_{imp}(r)$ produces short distance corrections
to the classical function $f_{cl}(r)$, while $\tilde f(r)$ produces large distance corrections
to $f_{cl}(r)$. Since one expects quantum corrections to be relevant rather at short
distance scales than at large distance scales, one can conclude that at this level
the ``improving solutions'' approach meets much better our physical expectations
than the ``solving improved actions'' approach. This seems to indicate that
the short distance corrections of   improved actions come from higher order, or even non-local operators.
This raises very interesting questions and possibilities for future investigations, in particular when working
with a ``solving improved actions'' approach.

\subsection*{3. Modified horizon structure, cosmic censorship, and temperature}

\begin{figure}[ht!]
  \centering
    \includegraphics[width=.6\textwidth]{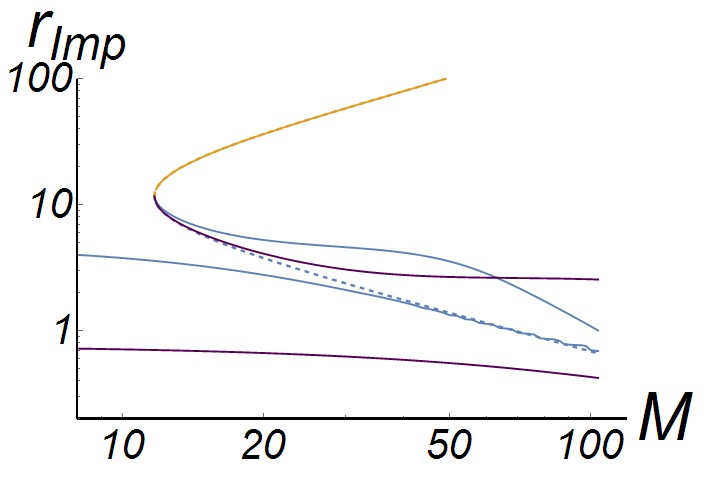}
\caption{\footnotesize{Improved Horizons, for dS black holes. The orange line corresponds to the external horizon and the blue lines are for internal horizons. Also we have purple lines, that correspond to internal horizons for $\xi=\xi_{sc}/3$. The dotted line is for the classical horizon. The values used are $G_0=1, Q_0=1, \alpha_0=1/137, k_0=0.01$}, and $\xi_{sc}$. For AdS black holes we have the same behavior.}
\label{fig:ImpHor}
\end{figure}

As discussed above, the RG-improvement induces for $\Lambda_0 \neq 0$ a structural change in the horizon
structure of the black hole. 
The most important feature is that a new internal horizon appears due to the improvement.
It is interesting to note that 
this new horizon exists even for masses below the critical mass value $M_1$. 
Since, one expects (and observes) that the improved solutions turns into the classical
solution for $\xi \rightarrow 0$ one has to understand how such a drastic feature, as the 
appearance of a new internal horizon, fits into this picture.
In order to understand this, one can plot the improved horizon structure
as a function of the black hole mass, just as it was done in the classical figure \ref{fig:horizons1}.
By varying the $\xi$ parameter in figure \ref{fig:ImpHor} one understands how the improved horizon structure,
with its additional horizon, fits to the classical horizon structure.
Due to the improvement, the classical inner horizon (dashed line) is split 
into two horizons, one with larger radius and the other one with smaller radius than the classical inner horizon.
For smaller and smaller values of $\xi$ the larger one of the internal horizons aligns with the classical inner horizon,
whereas the smaller additional horizon radius gets shifted towards $r_{(A)dS}\rightarrow 0$ recovering the classical behavior.
Still, as long as $\xi \neq 0$, this new inner horizon persists, even for masses below $M_1$.
The fact that there is a horizon below the critical mass is important for cosmic censorship hypothesis. 
In figure \ref{fig:Cosmic}, it was noted that there is a critical mass $M_c$ below which a naked singularity can appear.
From figure \ref{fig:ImpHor} one sees that this critical mass is pretty much unaffected by the improvement procedure.
However, the existence of a new internal horizon protects the improved black hole against the appearance
of a naked singularity, even for a arbitrarily small mass parameter $M_0$.
This means that, (for this choice of parameters \textendash scale setting), the weak cosmic censorship is fulfilled.

\begin{figure}[ht!]
  \centering
  \subfigure[\footnotesize{ }]{
    \includegraphics[width=.44\textwidth]{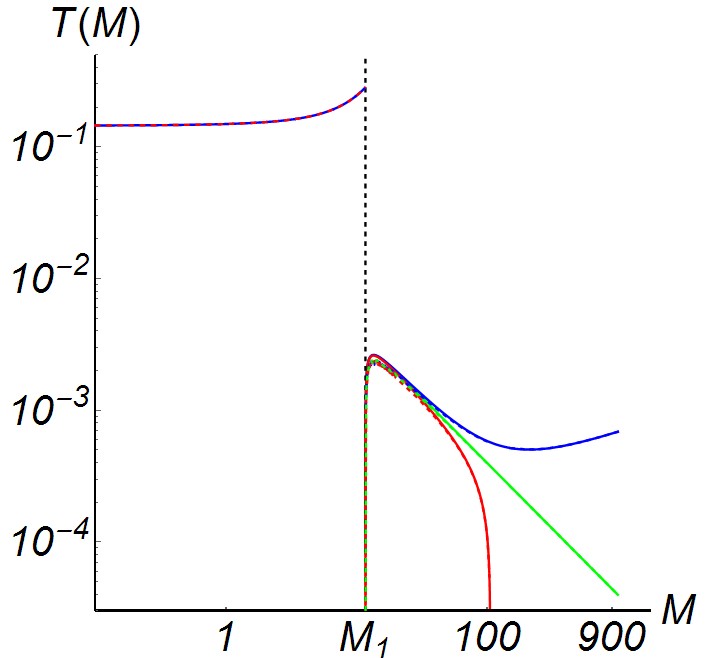}
  }\hspace{.1cm}
  \subfigure[\footnotesize{ }]{
    \includegraphics[width=.44\textwidth]{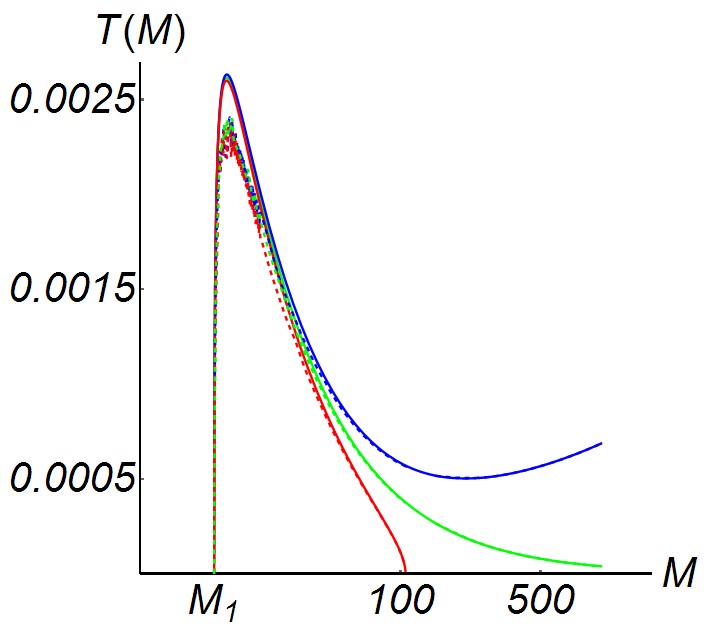}
  }
\caption{\footnotesize{Improved temperature as function of the mass parameter, evaluated in the external horizon for AdS (blue), dS (red) and $\Lambda_0=0$ (green) cases. Also, we plot in full lines the classical temperature \eqref{class_temp} evaluated in $r_2$ \eqref{r2}. The values chosen for evaluation are $G_0=1, Q_0=1, \alpha_0=1/137, k_0=0.01$}: \\
(a) shows the behavior for a large range of masses, where for $M<M_1$ the improve temperature is evaluated in the new internal improved horizon.\\
(b) is a zoom for masses $M>M_1$.}
\label{fig:tempImp}
\end{figure}

One of the most exciting facts about black holes is that they have a thermodynamic behavior,
which leads to a (quantum driven) evaporation process.
For describing this evaporation process for the classical black hole solution on uses the relation \eqref{class_temp},
which can be obtained from a comparison of the boundaries of this solution with
the asymptotic behavior of empty space-time.
However, the space-time of the improved
black hole solution has the same asymptotic behavior as the classical solution for large $r$.
Therefore, changes of the thermodynamic behavior due to the 
improved line element must be also encoded at the black hole horizon and not in the large $r$ asymptotics.
Therefore, it reasonable to use the relation \eqref{class_temp} for the description 
of the evaporation process of the improved black hole, where
the relation has then to be evaluated at the outer-inner horizon \citep{Hawking:1975,PhysRevD.15.2738}.
In figure \ref{fig:tempImp} the temperature of the improved black hole space-time is shown in dashed lines
as a function of the mass parameter $M_0$ in comparison to the classical temperature. The three different colors correspond to the dS, the AdS and
the $\Lambda_0=0$ case.
One observes that the improvement tends to lower the temperature of the improved black hole with respect to the classical black hole independent of $\Lambda_0$.
This difference in temperature is relatively small for the entire mass range.
Still one observes that it vanishes for large masses and is most pronounced for smaller masses. Notably, the
critical mass at which the temperature changes drastically is the same for the classical and the improved case.
This, can already be understood with figure \ref{fig:ImpHor}, since it is the mass at which two of the horizons merge,
which occurs to the same for the improved and the standard case.
However, while the temperature in the standard case drops to zero at this critical mass,
the improved solution provides a phase transition due to the additional inner horizon which
becomes visible at masses $M< M_{crit}$.
One observes that compared to this transition,
modification from the non-improved temperature are rather mild.

\subsection*{4. Modified mass, charge}

\begin{figure}[ht!]
  \centering
  \subfigure[\footnotesize{ }]{
    \includegraphics[width=.44\textwidth]{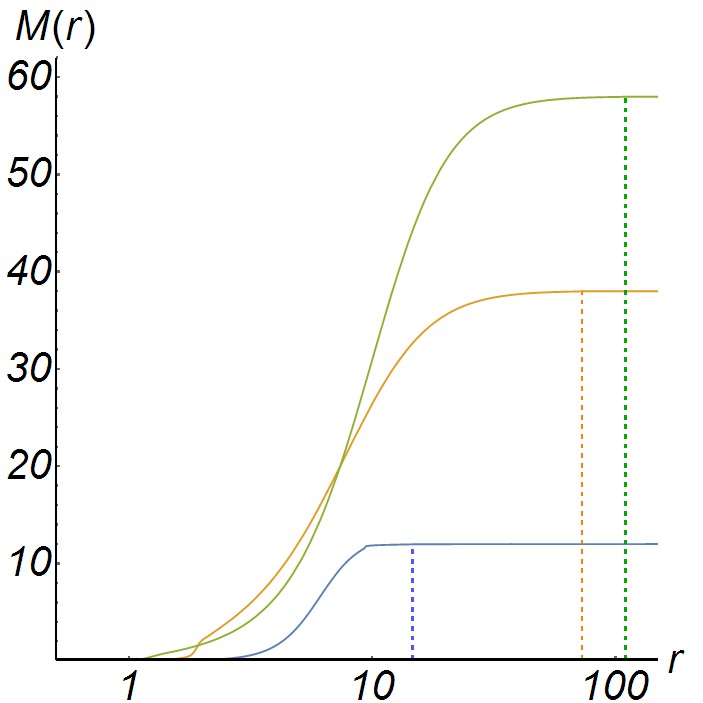}
  }\hspace{1cm}
  \subfigure[\footnotesize{ }]{
    \includegraphics[width=.44\textwidth]{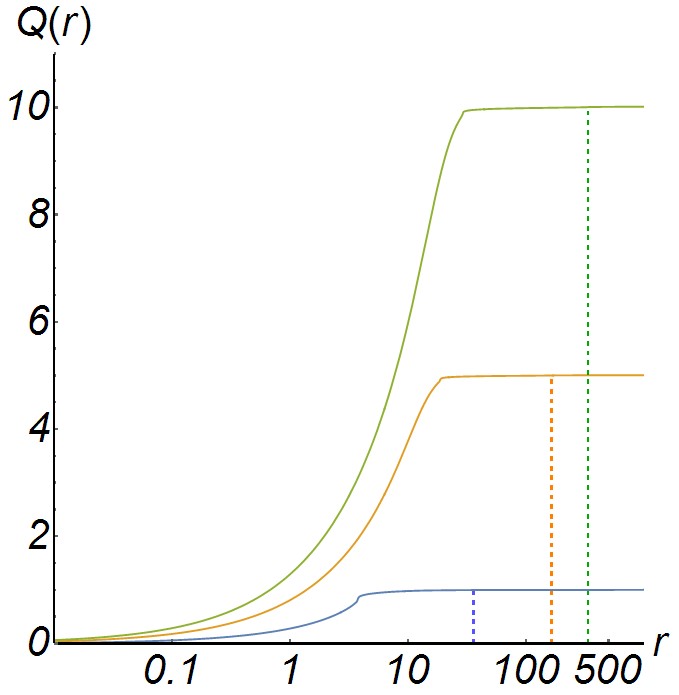}
  }
\caption{\footnotesize{Mass and charge as variables dependent of $r$ for AdS case: \\
(a) The different curves are for the mass values $M_0=\{12,~38,~58\}$ (from bottom to top), and fixed charge $Q_0=1$. \\
(b) The curves are for the charge values $Q_0=\{1,~5,~10\}$ (from bottom to top) with mass values $M_0=\{20,~100,~250\}$, respectively. The other parameters
 $\xi_{sc},~G_0=1,~\alpha_0=1/137,~k_0=0.01$}. 
 The dS case is basically identical since differences only would occur at extremely large radii.}
\label{fig:MQr}
\end{figure}

It is instructive to study how one would interpret the improved black hole solution
is one would not be aware of a possible scale dependence of the couplings.
In this case one would perform experiments at some radial scale $r$ and assuming constant couplings 
$G_0,\; \Lambda_0$, and $\alpha_0$. The result of such an experiment (say the study of sections of geodesics) would
then be fitted by the ``charges'' of the black hole. For astrophysical distances,
those charges would be basically the mass $M=M(r)$ and the electrical charge $Q^2=Q^2(r)$,
whereas the cosmological term with its corresponding ``charge'' $L=L(r)$ is largely irrelevant
at a range of smaller radii.
Taking equation \eqref{f_k} and redefining mass and charge in this sense of fitting the metric function one would write
\begin{equation}
f_k(r)=1-\frac{2G_0 M(r)}{r}+\frac{G_0 Q^2(r)}{\alpha_0 r^2}-\frac{1}{3}\Lambda_0 L(r) r^2\quad,
\end{equation}
with
\begin{equation}\label{M_r}
M(r)\equiv \frac{M_0 g(k)}{G_0 k^2(r)}\quad,
\end{equation}
and
\begin{equation}\label{Q_r}
Q(r)\equiv \frac{Q_0^2 \alpha_0}{G_0}\frac{g(k)}{\alpha(k)k^2(r)}\quad.
\end{equation}
This analysis allows to get a feeling on how such renormalization group effects would appear to an unprepared observer.
Figure \ref{fig:MQr} shows the radial dependence of the effective mass \eqref{M_r} and the effective charge \eqref{Q_r}.
One observes that for large radii, the effective masses and charges are basically identical to the
value of the classical parameters $M_0$ and $Q_0$. However, for smaller radii,
the effective mass and effective charge of the improved black hole
 are driven to smaller and smaller values, reflecting a screening effect of the scale dependence,
 that becomes effective for large scale $k$ and small scales $r$.

\section{Conclusions}
\label{SecConcl}

This paper studied the effects of renormalization group induced scale dependence
on the charged dS and AdS black hole solutions.
This is done by applying the ``improving classical solutions'' method suggested by Bonnano and Reuter \citep{Bonanno:2000ep}.
After introducing the scale dependent couplings in the classical solution \eqref{f_k}, the
arbitrary scale $k$ is set to physically relevant quantity by (\ref{k(r)}) resulting in the
line element of the improved black hole.
This line element shows agreement with the classical line element for large radial coordinates
and increasing deviations for small radial coordinates. 
The relative amount of corrections due to the renormalization
group effects is controlled by the $\xi$ parameter, defined in (\ref{k(r)}).
This parameter is implemented such that $\xi\rightarrow 0$ recovers the classical solution,
while increasing $\xi$ implies increasing relevance of the renormalization group effects.
As the name indicates, the improved black hole space-time is to be understood as
first short distance correction with respect to the classical result.
In this spirit the short distance improvement is formulated as an iterative process,
where an improved line element is used as bases for repeating the improvement process.
The crucial question then is, whether this procedure converges or not.
It is found that there exists a preferred choice for the parameter $\xi=\xi_{sc}$, 
which allows to get fastest ``convergence'' of the iterative procedure.
This choice agrees with the ``self-similar'' choice proposed in \citep{Koch:2013owa}
and it will be used as  benchmark for the numerical studies.

The significance and meaning of the differences between improved and
classical line elements 
and their dependence on the parameters 
is subsequently studied from various perspectives:
\begin{itemize}
\item Horizon structure: 
It turns out that the improved solutions have an additional internal horizon, 
which is present independent of the value of $M_0$
\item Cosmic censorship:
Classical charged black holes in (A)dS space have to invoke
the cosmic censorship in order to avoid the appearance of a naked singularity at $r\rightarrow 0$.
The improved charged black holes however, can count on an additional internal horizon,
which protects against the appearance of this naked singularity.
\item Temperature:
Like in the uncharged case \citep{Koch:2013owa},
the evaporation of the improved black hole does not stop at some remnant mass. 
But the improved charged black holes have the additional feature that at a mass corresponding to the classical critical
mass $M_{crit}$, the black hole experiences a zero order transition in the temperature. 
\item Effective mass and charge:
If one would not be aware of the scale dependence of the couplings one would associate
changes in the line element to changes in the charges, in particular $M=M(r)$ and $Q=Q(r)$.
One finds that those effective charges converge to the classical parameter $M_0$ and $Q_0$
at large radial parameters. 
However, at small radial parameters, the black hole appears to loose mass and charge, by virtue
of the renormalization group improvement. Such an effect can have interesting observational consequences \cite{Rodrigues:2015rya}.

\end{itemize}
The result on the cosmic censorship is quite remarkable and closely related
to the asymptotic behavior of the improved line element at small $r$.
In order to understand this limit, the appendix \ref{sec_ApA} studies
this effect analytically confirming the numerical observation of inverted asymptotics.\\

The work of B.K. was supported by Fondecyt 1120360 and Anillo Atlas Andino 10201.
The work of C.G. was supported by Conicyt 21130928, Conicyt 81140256 and proy. Fondecyt 1120360.
	
\begin{appendix}
\section{RG-Improved black hole solutions in the UV}\rq{7}
\label{sec_ApA}

An analytical expression for $k(r)$ not be obtained, unless for very extreme regimes, such as very small radial coordinates,
where one might dare to make an expansion in $r$
\begin{equation}\label{d(r)UV}
d(r)\simeq \frac{1}{2}\sqrt{\frac{\alpha_0}{G_0 Q_0^2}}r^2\left[1+\frac{2}{3}\frac{M_0\alpha_0^2}{Q_0^2}r+{\cal O}(r^2)\right].
\end{equation}
The cosmological constant $\Lambda_0$ does not appear in this expansion until order $r^6$. With this expression for $d(r)$, at first order in $r$, one can get an analytic result for $k(r)$ given by

\begin{equation}\label{k(r)UV}
k(r)\simeq 2Q_0\sqrt{\frac{G_0}{\alpha_0}}r^{-2}\xi.
\end{equation}
Taking the UV limit, $k\rightarrow \infty$, one must use (\ref{UV}) in (\ref{f_k}) obtaining
\begin{equation}
f_{*}(r)=1-\frac{2g_{*}M_0}{k^2r}+\frac{g_{*}Q_0^2}{\alpha_{*}k^2r^2}-\frac{1}{3}(\lambda_{*}k^2) r^2.
\end{equation}
Using the analytic expression for $k(r)$, one gets
\begin{equation}\label{f*}
f_{*}(r)=1-\frac{\alpha_0 g_{*} M_0}{2 G_0 \xi^2 Q_0^2} r^3+\frac{\alpha_0 g_{*} }{4 \alpha_{*} G_0 \xi^2}r^2-\frac{4 G_0 \lambda_{*} \xi ^2 Q_0^2}{3 \alpha_0 r^2}~.
\end{equation}
and the ``self-consistent'' improved function
\begin{equation}\label{f*sc}
f_{*,sc}(r)=1-\frac{2 \text{$\alpha_0$} g_{*} \lambda_{*}  M_0 r^3}{3 \text{G}_0 Q_0^2}+\frac{\text{$\alpha_0$} g_{*} \lambda_{*}  r^2}{3 \alpha_{*}  \text{G}_0}-\frac{\text{G}_0 Q_0^2}{\text{$\alpha_0$} r^2}~.
\end{equation}

One can see that this improved UV solution is analytically different from the classical solution (\ref{fr}). This is the origin of the differences of the complete improved and the classical solution that one can see in figure (\ref{fig:fimp}(a),(b)), because we can see this differences in the $r<M$ regime. Also, the fact that this UV approximation does not depend on the value of $\Lambda_0$, indicates that only in small distances is where differences are found.\\
Finally, with this difference of the classical and RG-Improved line element, one expects that some other structural modifications of the solution appears. One can compute the square of the Riemman-tensor of the UV RG-Improved solution (\ref{f*sc}), obtaining 
\begin{equation}\label{Rsc}
R_{\mu\nu\rho\sigma}R^{\mu\nu\rho\sigma}=\frac{8 \alpha_{0}^2 g_{*}^2 \lambda_{*}^2}{3 \alpha_{*} ^2 G_0^2}+\frac{304 \alpha_0^2 g_{*}^2 \lambda_{*} ^2 M_0^2 r^2}{9 G_0^2 Q_0^4}-\frac{160 \alpha_0 ^2 g_{*}^2 \lambda_{*}^2 M_0 r}{9 \alpha_{*}  G_0^2 Q_0^2}+\frac{64 g_{*} \lambda_{*}  M_0}{3 r^3}+\frac{56 G_0^2 Q_0^4}{\alpha_0^2 r^8}~.
\end{equation}
where one can see that the spatial singularity at the origin $r=0$ remains. This mean that RG-Improve process does not resolve the singular behavior of the classical solution in contrast of the calculations made for Reuter and Bonanno \cite{Bonanno:1998ye} for Schwarzschild black holes where they remove the spatial singularity at the origin with the RG-Improvement.

\end{appendix}\newpage

\bibliography{bibliography}

\end{document}